# Relaxation dynamics of poly(ethylene oxide)

Peter Lunkenheimer* &  Alois Loidl

Experimental Physics V, Center for Electronic Correlations and Magnetism, University of Augsburg, 86135 Augsburg, Germany

*Corresponding author; e-mail: peter.lunkenheimer@physik.uni-augsburg.de

**Poly(ethylene oxide) is an important polymer with many applications, e.g., as solid-state electrolyte in batteries. Its relaxation dynamics, characterizing its molecular and submolecular motions, which is relevant for many of these applications, was investigated numerous times, mostly employing dielectric spectroscopy. However, the various dynamic processes revealed by these studies were interpreted in conflicting ways and even their nomenclature in literature is highly inconsistent. Here we present the results of a detailed investigation of this polymer employing dielectric spectroscopy covering a relatively broad frequency and temperature range. We clearly detect four intrinsic relaxation processes. The slowest one most likely represents a so-called normal mode, reflecting global motions of the polymer chains, an interpretation that was not considered in previous works. The second process can be unequivocally identified with the segmental $\alpha$ relaxation, which governs glassy freezing and the glass transition. The third, only rarely detected process corresponds to the Johari-Goldstein relaxation of poly(ethylene oxide), widely overlooked in previous studies. The fourth and fastest process is unrelated to the supercooled and glassy state of this polymer and probably due to local, intramolecular motions.**

Investigations of the dielectric properties and relaxation dynamics of polymers not only are relevant for application, e.g., as insulator materials, but are also highly important from a more fundamental point of view: They can provide valuable information about the different types of dynamics on a molecular level, which govern, e.g., the glass transition or the conductivity of ion-doped polymers, considered as solid-state electrolytes. Poly(ethylene oxide) (PEO), whose low-molecular-weight variants are also termed poly(ethylene glycol), has innumerous applications in technology, medicine, and biology. In recent years, interest in this material was renewed due to its good salt-solvating ability, leading to high ionic conductivity, which makes this polymer one of the most studied polymers for the future use as electrolyte in lithium-ion batteries[1,2]. There are numerous publications reporting the application of dielectric spectroscopy on PEO (e.g., refs. 3-13). However, usually only certain, mostly quite limited temperature and frequency ranges are covered by these different studies and the molecular weight of the investigated samples often varies considerably. Surveying all these works reveals that four intrinsic dipolar relaxation processes seem to exist in this polymer, although this fact is not becoming clear in any of these papers. In fact, the slowest process is often assumed to be inaccessible by dielectric spectroscopy[6,10,11], but it seems that in two works[4,12] (to our knowledge) it was indeed detected without noting the significance of this finding. Moreover, some confusion arises because the nomenclature of the different processes often is inconsistent. Therefore, it is difficult to gain a comprehensive picture of the overall relaxational dynamics in this polymer from the current literature. During our dielectric investigation of PEO with added lithium bis(trifluoromethanesulfonyl)imde (LiTFSI)[14], as a reference material we also investigated pure PEO in a broad frequency and temperature range. Here we report these data, which cover all four dynamic processes in the same sample material having the same molecular weight of $10^5$ g/mol. For two of these processes, we provide interpretations that were not considered in previous dielectric studies of this polymer. It should be noted that PEO tends to be of semicrystalline nature with coexisting crystalline and amorphous phases[8,12,15]. Nevertheless, its relaxational dynamics mostly exhibits the typical signatures of amorphous polymers and, thus, is believed to primarily arise from the disordered regions[7,8,11]. However, a partial influence of the crystalline regions on the relaxational behavior of PEO, e.g., due to the constraint of dipolar motions imposed by the crystallites, was also considered[3,8,11].

## Dielectric spectra and identification of four relaxation processes

Figure 1 shows spectra of the complex permittivity $\varepsilon^* = \varepsilon' - i\varepsilon''$, namely the dielectric constant $\varepsilon'$ (a) and loss $\varepsilon''$ (b), of PEO for various temperatures between 158 and 281 K. At high temperatures, the real part reveals a prominent steplike decrease with increasing frequency, which shifts to lower frequencies with decreasing temperature. These are the typical signatures of a dipolar relaxation process[16,17], and we term it $\alpha$ relaxation in the following (we come back to the nomenclature later). Its relaxation strength can be estimated from the step amplitude (from about 3 to 2) to be of order one. In the dielectric loss (Fig. 1b), this process shows up as a peak, approximately located at the points of inflection of the $\varepsilon'$ spectra. At low temperatures, $T \leq 224$ K, the high-frequency tails of the $\alpha$-relaxation steps in Fig. 1a appear rather smeared-out. Moreover, at the lowest temperatures, when the main step has shifted out of the frequency window, the $\varepsilon'$ spectra reveal the faint indication of another step or shoulder, pointing to a further process. The existence of such an additional relaxation, faster than the $\alpha$ process, is indeed clearly confirmed by the loss spectra (Fig. 1b). With decreasing temperature, starting from 224 K, beyond the main peak first a weak hump shows up (e.g., at about 1 MHz for the 224 K curve) which then evolves into a clearly discernible peak. It shifts to low frequencies and finally shows up at about 200 Hz for the lowest temperature of 158 K. We term this process $\gamma$ relaxation. Moreover, for 212 K, between the $\gamma$ relaxation and the $\alpha$ relaxation (only revealed by its high-frequency flank at this temperature), a shoulder appears, developing into a peak upon cooling. Consequently, for example at 191 K the loss spectrum exhibits two peaks, the right one due to the $\gamma$ relaxation and the left one due to this additional process which we term $\beta$ relaxation. The amplitude of the latter is too small to lead to any discernible step in the real part (Fig. 1a).

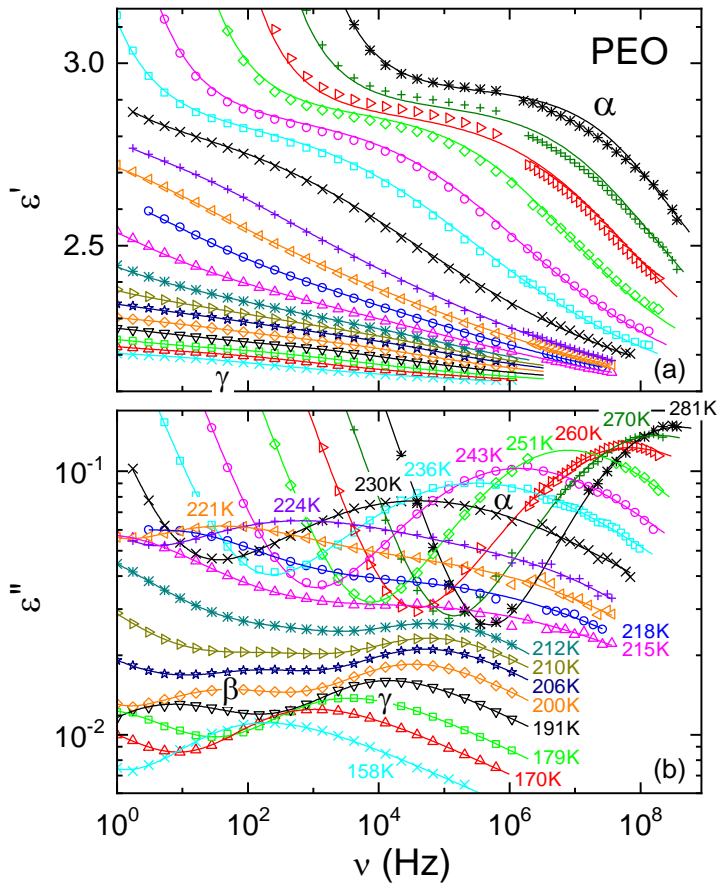

**Figure 1.** Spectra of $\varepsilon'$ and $\varepsilon''$ of PEO for temperatures 158–281 K as indicated in (b). The solid lines are fits, simultaneously performed for both quantities, as described in the text.

Thus, overall, three separate relaxation processes can be clearly identified in Fig. 1. Moreover, at frequencies below the $\alpha$ relaxation, indications of a fourth contribution become obvious: There both $\varepsilon'$ and $\varepsilon''$ exhibit an additional increase with decreasing frequency, finally leaving the frames of the corresponding graphs, whose ranges were chosen to clearly reveal the $\alpha$, $\beta$, and $\gamma$ relaxations treated above. For the relevant temperatures, $T \geq 230$ K, Figure 2 shows the same spectra as in Fig. 1 (using identical symbols to facilitate comparison), but with the ordinate extending up to higher values, beyond $10^3$. This figure also presents spectra for additional temperatures up to 324 K, not included in Fig. 1 because, above 281 K, the $\alpha$-relaxation peak has shifted out of the frequency window. It reveals that the mentioned low-frequency increase of $\varepsilon'$ and $\varepsilon''$, whose onset is seen in Fig. 1, extends over many decades. In both quantities, it exhibits a shoulder (e.g., at around 1 kHz for the highest temperature) which points to a fourth relaxation in PEO, located at lower frequencies than the $\alpha$ process. Its relaxation strength, estimated from the step-height in $\varepsilon'(\nu)$ (Fig. 2a), is of the order 10, significantly larger than that of the $\alpha$ relaxation, indicating larger dipole moments of the entities causing this process. It probably represents a normal-mode relaxation as will be discussed in more detail below. In the following, we will term it $\alpha'$ relaxation. In the loss spectra, Fig. 2b, this process is not disclosed by clearly visible peaks. Instead, due to the superposition by another contribution at lower frequencies, only the high-frequency flanks of the $\alpha$ peaks and a shoulder show up. In the $\varepsilon'$ spectra, at the lowest frequencies and highest temperatures this additional low-frequency process leads to very large values beyond $10^3$. This is unrealistically high for a dipolar relaxation. As discussed in detail in ref. 18 such high values can be found for ferroelectrics (excluded here), non-intrinsic Maxwell-Wagner relaxation (e.g., due to blocking electrodes at the sample-surface interface of ionic conductors), or by charge transport via hopping conductivity.

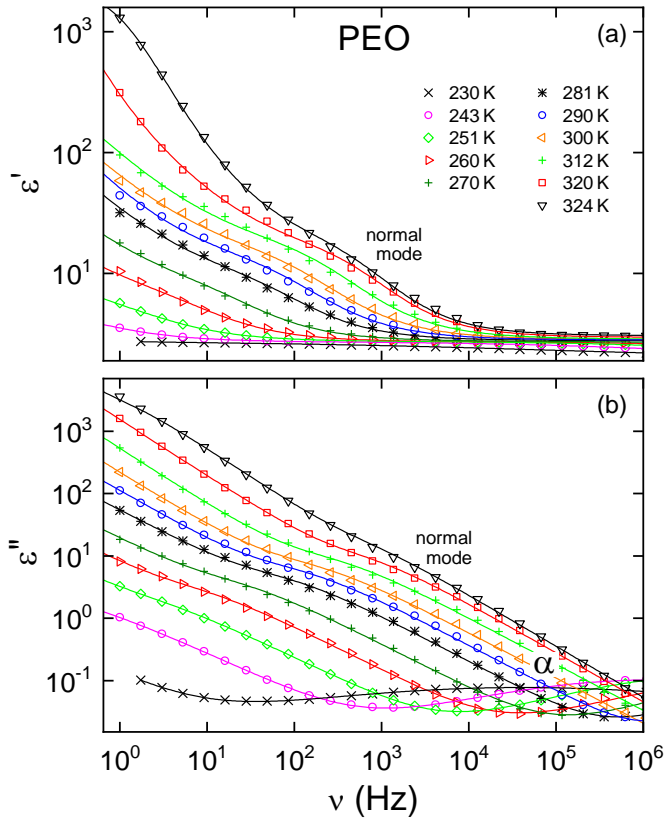

**Figure 2.** Spectra of $\varepsilon'$ and $\varepsilon''$ of PEO for temperatures 230–324 K as indicated in the legend. The solid lines are fits, simultaneously performed for both quantities, as described in the text.

Overall, we have identified the signatures of four dipolar relaxation processes in the spectra of Figs. 1 and 2. It should be noted that in none of these spectra, taken at different temperatures, all four processes show up simultaneously. For example, at 210 K in Fig. 1b the $\alpha$, $\beta$, and $\gamma$ relaxations can be identified ($\alpha$: only right flank visible; $\beta$: strongly superimposed by $\alpha$ and $\gamma$) but the $\alpha'$ mode is outside the lower boundary of the frequency window. On the other hand, 230 K is the lowest temperature where the right flank of the $\alpha'$ peak is just seen (Fig. 2b), but even here $\beta$ and $\gamma$ are already located at too high frequencies (or merged with $\alpha$) to be detectable (Fig. 1b). All this makes it difficult to detect the simultaneous existence of four processes in this polymer, which may be one of the reasons why this fact was overlooked by earlier dielectric investigations of PEO.

**Analysis of the spectra**

To learn more about these relaxations and to disclose their most important parameter, the relaxation time, we have fitted the spectra (solid lines in Figs. 1 and 2). The most common fit function for dipolar relaxations is the empirical Havriliak-Negami (HN) formula[19]:

$$\varepsilon^* = \varepsilon_\infty + \frac{\varepsilon_s - \varepsilon_\infty}{[1 + (i\omega\tau)^{1-\alpha}]^\beta} \qquad (1)$$

Here $\varepsilon_\infty$ is the high-frequency limit of $\varepsilon'$, $\varepsilon_s$ denotes the static dielectric constant, and $\tau$ is the relaxation time. $\alpha$ and $\beta$ are width parameters, which determine the symmetric and asymmetric loss-peak broadening, respectively. For $\beta = 1$, the peaks become symmetric, and Eq. (1) is identical to the Cole-Cole (CC) equation[20]. For $\alpha = 0$, Eq. (1) corresponds to the often-employed Cole-Davidson (CD) formula[21]. The four detected relaxations were fitted with the sum of four HN equations, however, trying to minimize the number of parameters by setting either $\beta = 1$ or $\alpha = 0$, if possible. We find that for the $\alpha'$ and $\beta$ relaxation, the CC function ($\beta = 1$) is sufficient for all temperatures where these processes are detected. For the $\alpha$ process, the full HN function had to be used, except for the lowest temperatures, where its asymmetry cannot be detected because its low-frequency flank is outside the frequency window. There, the CC function could be used. For the $\gamma$ relaxation, the HN function was employed for $T \leq 212$ K, while at higher temperatures, the CC function was sufficient. A close inspection of Fig. 1b reveals that this crossover from HN to CC may be an artefact due to the lack of data at $\nu > 1$ MHz below 212 K, as there the relatively low loss values were outside the resolution limit of the high-frequency device. Anyway, as the $\gamma$ peaks are well pronounced at low temperatures, the relaxation times obtained from the fits are reliable.

The mentioned additional low-frequency increase of $\varepsilon'$ and $\varepsilon''$ below the $\alpha'$ relaxation was mainly accounted for by additional ac- and dc-conductivity contributions. In the permittivity, due to the general relation $\varepsilon^* = \sigma^*/(i\varepsilon_0\omega)$[22], where $\sigma^* = \sigma' + i\sigma''$ is the complex conductivity, dc conductivity causes a term $\varepsilon^*_{dc} = \sigma_{dc}/(i\varepsilon_0\omega)$ (here $\varepsilon_0$ is the permittivity of free space and $\omega = 2\pi\nu$). It leads to $\varepsilon''_{dc} = \sigma_{dc}/\varepsilon_0\omega$ and no contribution in $\varepsilon'(\nu)$. Ac conductivity can be modeled by the so-called universal dielectric response (UDR)[23], given by $\sigma^*_{ac} \propto (i\omega)^s$, with $s \leq 1$. It is usually assumed to arise from hopping conductivity of localized charge carriers and detected in various types of disordered matter and ionic conductors[23,24,25,26]. In the permittivity, it leads to $\varepsilon^*_{ac} \propto (i\omega)^{s-1}$, resulting in a $\nu^{s-1}$ power law in both, $\varepsilon'(\nu)$ and $\varepsilon''(\nu)$. This accounts for the low-frequency increases observed in Figs. 2a and b at high temperatures, which partly obscure the $\alpha'$ relaxation. Finally, for the two highest temperatures in Fig. 2 an RC equivalent circuit in series to the sample had to be assumed to model a small contribution from blocking electrodes[18,27]. It seems reasonable that in PEO, which is prone to ionic charge transport, these low-frequency spectral contributions arise from translational motions of residual amounts of ions. Calculating $\sigma'$ from $\varepsilon''$ in Fig. 2b, even for the highest temperature a rather small low-frequency conductivity of order $10^{-9}$ $\Omega^{-1}$cm$^{-1}$ is obtained showing that, in principle, this is a small effect which, however, becomes dominant in the permittivity at high temperatures due to the mentioned $\nu^{s-1}$ power law. As it is likely caused by impurity-induced ions, we refrain from discussing it in more detail here.

The lines in Figs. 1 and 2 are fits with the contributions discussed in the preceding paragraphs, simultaneously performed for $\varepsilon'$ and $\varepsilon''$. Overall, a good description of the experimental data is reached in this way. One may criticize that the fits make use of an excessive number of fit parameters. However, we want to point out that, in each fit of the individual spectra, only few of the above-discussed contributions had to be simultaneously employed. For example, at high temperatures, only the conductivity and $\alpha$ relaxation is needed, while at low temperatures only the $\beta$ and $\gamma$ relaxations show up, which limits the number of fit parameters

**Relaxation times**

*$\alpha$ relaxation*

The HN function implies a broadening of the loss peaks compared to the Debye function (Eq. (1) with $\alpha = 0$ and $\beta = 1$)[16,19]. This is a common phenomenon in disordered matter, usually ascribed to a distribution of relaxation times[28,29,30]. From the fit parameter $\tau$ in Eq. (1), an average relaxation time $\langle\tau\rangle$ should be calculated and used for the analysis. For the CC function, $\langle\tau\rangle = \tau$ and for the CD function $\langle\tau\rangle = \beta\tau$. For the HN function, $\langle\tau\rangle$ is not defined[31], and we used the peak frequency $\nu_p$, calculated[32] from the parameters of Eq. (1), to estimate $\langle\tau\rangle$ via $\langle\tau\rangle \approx 1/(2\pi\nu_p)$. The average relaxation times of all four processes, as obtained from the fits, are shown by the closed symbols in Fig. 3. Within this Arrhenius representation, the $\tau(1000/T)$ data of the $\alpha'$, $\alpha$, and $\beta$ relaxation all can be reasonably fitted by straight lines. This evidences thermally activated temperature dependence, $\langle\tau\rangle \propto \exp[E/(k_B T)]$, where $E$ is the energy barrier and $k_B$ the Boltzmann constant. The linear fits lead to $E$ values of 0.53 ($\alpha'$), 0.69 ($\beta$), and 0.36 eV ($\gamma$). In contrast, the $\alpha$-relaxation time (closed red squares in Fig. 3) reveals significant deviations from Arrhenius behavior. It can be fitted by the empirical Vogel-Fulcher-Tammann (VFT) equation[33,34,35,36],

$$\langle\tau\rangle = \tau_0 \exp\left[\frac{D_\tau T_{VF}}{T - T_{VF}}\right], \quad (2)$$

commonly applied to supercooled liquids[17,37]. Here $\tau_0$ can be regarded as an inverse attempt frequency, $D$ is the so-called strength parameter[36,37] characterizing the deviations from Arrhenius behavior, and $T_{VF}$ is the Vogel-Fulcher temperature. The red line in Fig. 3 is a fit with Eq. (2) of the present $\alpha$-relaxation data (closed squares), combined with those from Porter *et al.* (open squares), which nicely complement our data at very small $\tau$ values. The experimental data can be reasonably fitted in this way, leading to $\tau_0 = 5.17\times10^{-14}$ s, $D = 4.81$, and $T_{VF} = 185$ K. Non-Arrhenius temperature dependence is a typical property of the $\alpha$-relaxation dynamics of supercooled materials, universally found, e.g., in molecular liquids, metallic glasses, and polymers[17,36,37]. In the latter, the $\alpha$ relaxation is ascribed to segmental motions, which also govern the viscosity and whose freezing at low temperatures determines the glass transition[38]. Using the often-assumed condition $\langle\tau\rangle(T_g) \approx 100$ s to define the glass-transition temperature $T_g$, from an extrapolation of the red line in Fig. 3 we arrive at $T_g \approx 211$ K. In literature, for PEO with a molecular weight of order $10^5$ g/mol, various $T_g$ values were reported, e.g., 217 K (ref. 15), 219 K (ref. 39), or 226 K (ref. 40). Considering that these values were obtained using different experimental methods and definitions of $T_g$, they are compatible with the present result, further corroborating the correct assignment of the $\alpha$ relaxation.

This relaxation process was previously detected in several dielectric investigations[3-13] (often, but not always, termed $\beta$ relaxation) and $\tau$ values derived from these works are shown in Fig. 3. They are mostly based on the reported or read-off loss-peak positions in $\varepsilon''(T)$ (refs. 3,5,8,9) and $\varepsilon''(\nu)$ plots[3,4,7,11] and were partly derived from fits of the loss spectra[11,12]. Moreover, results from mechanical measurements are also included.[3,41] The general temperature dependence of these literature data, including the characteristic non-Arrhenius temperature dependence, agrees with that of the present ones, shown by the closed red squares. However, the absolute values partly deviate considerably. This mostly reflects the different ways of determining and defining $\tau$ and the variation in molecular weight of the investigated samples, which is indicated in the legend of Fig. 3. It should be noted that in none of these works simultaneous fits of the real and imaginary parts of the permittivity were performed, a procedure that enhances the precision of the present results. In refs. 3-13, this relaxation was interpreted and named in various ways, which we will not discuss in detail here. We think, the above discussion clearly confirms that it represents the segmental relaxation of PEO governing its glass transition[5,6,10,11] and thus, following common practice in polymer[38] and glass physics[17,37], should be termed $\alpha$ relaxation. Finally, we want to mention that this $\alpha$ relaxation of PEO and, consequently, its $T_g$ were reported to significantly depend on the molecular weight and the degree of crystallinity of the investigated samples[3,8,9,40]. In principle, segmental dynamics should be rather independent of the polymer chain length[38]. However, in PEO crystallinity and molecular weight seem to be correlated[3,6,42] and the dependence of the $\alpha$ relaxation on the latter then can be assumed to be due to constraints imposed on the segmental dynamics by the partially crystalline environment[3,6,8,11]. In single-crystalline PEO, as expected the $\alpha$ relaxation is completely absent[5].

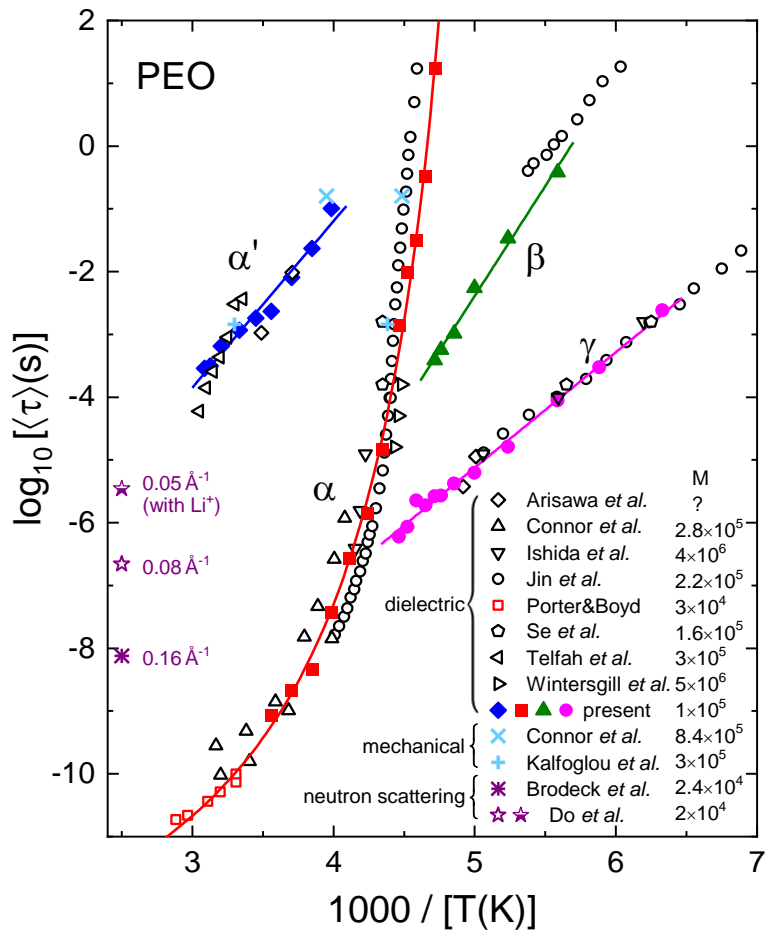

**Figure 3.** Average relaxation times of the four processes detected in PEO as derived from the fits of the permittivity spectra in Figs. 1 and 2, plotted within an Arrhenius representation (filled symbols). The open symbols show $\tau$ data as reported from dielectric investigations in literature[3,4,5,7,8,9,11,12] (the molecular weights of the measured samples are indicated in the legend; $\tau$ from ref. 3 was calculated from the reported loss-peak positions in both $\varepsilon''(T)$ and $\varepsilon''(\nu)$ plots). The crosses (+ and ×) show $\tau$ from mechanical measurements[3,41]. The different stars close to the left ordinate are results at 400 K from neutron scattering for different $q$ values as indicated in the figure[14,45]. The uppermost star was measured for PEO with added Li ions (with an EO:Li ratio of 10/1)[14]. The lines are fits of the present data with the VFT equation, Eq. (2), for the $\alpha$ relaxation and with the Arrhenius law for the other relaxations.

### $\alpha'$ relaxation

The closed blue diamonds in Fig. 3 indicate the relaxation times of the $\alpha'$ process as obtained from the fits of the dielectric spectra in Fig. 2. It is many decades slower than the $\alpha$ relaxation. As revealed by the additional data points from literature included in Fig. 3 in this region, this process was previously detected by mechanical[3,41] and also by dielectric investigations[4,12]. In these works, it was interpreted as segmental dynamics[4,12] or motions related to crystal phases or crystalline boundary regions[3,41]. (It should be noted that this process was termed $\alpha$ relaxation in various earlier works.) The $\tau(T)$ data reported in ref. 12 in the rather limited temperature range of 298–328 K (left-pointed triangles in Fig. 3) exhibit stronger temperature dependence than the other data on the $\alpha'$ relaxation, for which we currently have no explanation. In ref. 13, the detection of a normal mode in PEO was claimed based on dielectric-loss data. However, as only a power law to the left of the $\alpha$ process and no significant shoulder in $\varepsilon'(\nu)$ or $\varepsilon''(\nu)$ was detected (cf. Fig. 2), in our view the significance of this claim is very limited. No relaxation times of the normal mode were provided in this work. Here we also should mention a dielectric investigation of polystyrene/PEO composites, where two relaxation processes slower than the $\alpha$ one were found and assumed to reflect intrinsic PEO dynamics[43]. As these samples were highly heterogeneous and only contained 5 wt% of PEO, the relevance of these results for bulk PEO, however, may be questioned, and we do not include the corresponding data points in Fig. 3.

Concerning the microscopic origin of the $\alpha'$ relaxation, one should be aware that a relatively slow dynamic process in PEO (compared to its $\alpha$ relaxation) was previously reported by several studies using neutron-scattering and/or molecular-dynamics simulations[44,45,46,47]. By a comparison of the experimental and simulation results, this process was clearly identified as a Rouse mode[45,46]. In accord with theoretical predictions, its relaxation time was found to increase strongly with decreasing momentum transfer $q$ of the scattering experiment. Moreover, in ref. 14, by neutron-scattering experiments on PEO with added lithium salt, a Rouse mode (denoted normal-mode there) also was clearly identified. In simple terms, this type of polymer dynamics arises from motions of the whole polymer chain, involving a variation of its end-to-end vector, in contrast to the $\alpha$ relaxation governing the glass transition and viscous flow, which usually is regarded as a segmental motion, taking place on much shorter spatial scale[38,47]. Interestingly, Rouse modes in principle can also be detected via dielectric spectroscopy, where they are usually termed normal-mode relaxations[38,47]. Could the $\alpha'$ relaxation found in the present work be such a normal mode? The stars close to the left ordinate in Fig. 3 present relaxation times as obtained by neutron scattering at 400 K for different $q$ values[14,42] (as we did not find results for pure PEO at very small $q$ values, we also include one data point from the mentioned study of lithium-doped PEO[14]). Formally, dielectric spectroscopy can be regarded as $q = 0$ experiment and Fig. 3 reveals that these $\tau$ values from neutron scattering come close to those of the dielectrically detected $\alpha'$ process for small $q$. This indeed indicates that the slowest relaxation process in PEO could be a normal mode. One should be aware that, in principle, normal modes in dielectric spectra are only expected for so-called type A polymers[48,49]. They are characterized by dipolar moments along the chain backbone, which is the case for PEO, but if these dipoles cancel each other out like in PEO, the polymer is not considered as "intrinsic" type A[49]. However, in ref. 49 it was pointed out that, even if the latter condition is not fulfilled, a material may behave as a type-A polymer if it has a helical structure, which indeed is the case for PEO[6].

### $\beta$ and $\gamma$ relaxations

Finally, there are two secondary relaxation process in PEO, whose $\tau(T)$ follows Arrhenius behavior as mentioned above (Fig. 3). We term them $\beta$ (at lower frequencies) and $\gamma$ (at higher frequencies). The $\gamma$ relaxation was detected in various earlier works and the reported relaxation times reasonably match those from the present work (Fig. 3). It was ascribed to local segmental dynamics[8], local twisting motions in the main chains[5,11], or to "local rotating and twisting polar side"[13]. The $\beta$ relaxation was first discovered in ref. 11 (denoted $\gamma'$ there) and ascribed to "the motion of PEO segments in the transition region between PEO lamellae and disordered interlamellar amorphous segments". While the $\tau(T)$ results for this process from ref. 11 do not perfectly match those from the present work (Fig. 3), it is clear that the same dynamics is detected.

Concerning the microscopic origin of these two secondary relaxations, we want to remark that one of them most likely is of Johari-Goldstein (JG) type[50]. The occurrence of JG relaxations seems to be a universal property of all classes of glass-forming matter, including polymers[38,51,52]. While their microscopic explanation still is controversial, it is clear that they do not arise from intramolecular degrees of freedom, e.g., dipolar-side-group dynamics in certain polymers, which can be excluded for PEO anyway. Instead, JG relaxations were proposed to be due to accelerated relaxation dynamics within so-called islands of mobility[50] or motions within fine structures of the energy potential experienced by the relaxing entities[53,54]. A universal JG relaxation is also predicted by the coupling model treating the dynamics of supercooled liquids and glasses[52,55]. In a very recent work[56], the $\beta$ relaxation of PEO, as detected in dielectric measurements of the pure polymer in ref. 11 (termed $\gamma$ there) and of polymer blends in ref. 57 (termed $\beta'$ or $\gamma'$), was proposed to be of JG type. Notably, the $\gamma$ relaxation discussed above was also detected in single-crystalline PEO[5]. This excludes that it is of JG type, further corroborating that indeed the $\beta$ relaxation represents the JG relaxation of PEO.

**Concluding remarks**

In summary, we have performed a detailed investigation of PEO using dielectric spectroscopy. Our results reveal all four intrinsic relaxation processes, for which we propose a consistent nomenclature, based on their properties. We tentatively assign the slowest process of PEO (termed $\alpha'$) to a normal mode relaxation, reflecting the global chain dynamics. This is mainly based on a comparison with neutron-scattering results, clearly identifying a Rouse mode. The main dynamics detected at higher frequencies (or shorter times) reveals the typical characteristics of supercooled liquids and clearly represents the $\alpha$ relaxation of PEO, governing its glassy freezing. The rarely detected $\beta$ relaxation at higher frequencies is identified with the JG relaxation of this polymer, universally expected for all classes of disordered matter. Finally, the fastest relaxation, denoted as $\gamma$, is not associated with the supercooled or glassy state of PEO and most likely due to local, intramolecular motions.

**Methods**

The samples with molecular weight of $10^5$ g/mol were purchased from Sigma-Aldrich. The slightly heated sample material was filled into parallel-plate capacitors with plate distances of the order of 0.1 mm, ensured by inserted glass-fiber spacers. The low-frequency measurements at 1 Hz – 1 MHz were performed using a frequency response analyzer (Novocontrol alpha-analyzer). At higher frequencies, $\nu > 1$ MHz, a coaxial reflectometric technique was used employing an Agilent 4991A impedance analyzer[58]. For cooling and heating, the sample was put into a $N_2$-gas cryostat (Novocontrol Quatro). The high-frequency measurements were only performed below room temperature.

**Data availability**

The data of this study are available from the corresponding author upon reasonable request.

**Acknowledgements**

We thank F. Humann for performing the dielectric measurements.

**Author contributions**

P.L. conceived and designed the study, analyzed the data, and wrote the manuscript. A.L. critically reviewed and edited the manuscript and provided funding support.

**Competing interests**

The authors declare no competing interests.

**Additional information**

Correspondence and requests for materials should be addressed to P.L.

**References**
1. Xue, Z., He, D. & Xie, X. Poly(ethylene oxide)-based electrolytes for lithium ion batteries. *J. Mater. Chem. A* **3**, 19218–19253 (2015).
2. Bocharova, V. & Sokolov, A. P. Perspectives for polymer electrolytes: a view from fundamentals of ionic conductivity. *Macromolecules* **53**, 4141–4157 (2020).
3. Connor, T. M., Read, B. E. & Williams, G. The Dielectric, Dynamic, Mechanical and Nuclear Resonance Properties of Polyethylene Oxide as a Function of Molecular Weight. *J. Appl. Chem.* **14**, 74−81 (1964).
4. Arisawa, K., Tsuge, K. & Wada, Y. Dielectric relaxations in polyoxymethylene and polyethylene oxide. *Jpn. J. Appl. Phys.* **4**, 138–147 (1965).
5. Ishida, Y., Matsuo, M. & Takayanagi, M. Dielectric behavior of single crystals of poly(ethylene oxide). J. *Polym. Sci. Part B Polym. Lett.* **3**, 321–324 (1965).
6. McCrum, N. G., Read, B. E. & Williams, G. *Anelastic and Dielectric Effects in Polymeric Solids* (Wiley, New York, 1967).
7. Porter, C. H. & Boyd, R. H. A dielectric study of the effects of melting on molecular relaxation in poly (ethylene oxide) and polyoxymethylene. *Macromolecules* **4**, 589–594 (1971).
8. Se, K., Adachi, K. & Kotaka, T. Phase behavior of binary and ternary fluoropolymer (PVDF-HFP) solutions for single-ion conductors. *Polym. J.*, **13**, 1009–1017 (1981).
9. Wintersgill, M. C., Fontanella, J. J., Welcher, P. J. & Andeen, C. G. High-pressure and molecular weight variation of electrical relaxation in poly(ethylene oxide). *J. Appl. Phys.* **58**, 2875–2878 (1985).
10. Fanggao, C. *et al*. Temperature and frequency dependencies of the complex dielectric constant of poly(ethylene oxide) under hydrostatic pressure. *J. Polym. Sci. Part B Polym. Phys.* **34**, 425–433 (1996).
11. Jin, X., Zhang, S. H. & Runt, J. Observation of a fast dielectric relaxation in semi-crystalline poly(ethylene oxide). *Polymer* **43**, 6247– 6254 (2002).
12. Telfah, A. *et al*. HR MAS NMR, Dielectric Impedance and XRD Characterization of Polyethylene Oxide Films for Structural Phase Transitions. *Physica B* **646**, 414353 (2022).
13. Bozoglu, D., Yakut, S., Ulutas, K. & Deger, D. Comparison of the effect of free volume fraction on the dielectric properties of polyethylene oxide bulk and thin film. *J. Non-Cryst. Solids* **625**, 122750 (2024).
14. Do, C. *et al*. Li$^+$ Transport in Poly(Ethylene Oxide) Based Electrolytes: Neutron Scattering, Dielectric Spectroscopy, and Molecular Dynamics Simulations. *Phys. Rev. Lett.* **111**, 018301 (2013).
15. Vrandečić, N. S., Erceg, M., Jakić, M. & Klarić, I. Kinetic analysis of thermal degradation of poly(ethylene glycol) and poly(ethylene oxide)s of different molecular weight. *Thermochim. Acta* **498**, 71–80 (2010).
16. Schönhals, A. & Kremer, F. Analysis of dielectric spectra, in *Broadband Dielectric Spectroscopy* (ed. Kremer, F. & Schönhals A.) 59–98 (Springer, Berlin, 2002).
17. Lunkenheimer, P. & Loidl, A. Glassy Dynamics: From Millihertz to Terahertz, in *The Scaling of Relaxation Processes* (ed. Kremer. F. & Loidl. A.) 23–59 (Springer, Cham, 2018).
18. Lunkenheimer, P. *et al*. Colossal dielectric constants in transition-metal oxides. *Eur. Phys. J. Spec. Top.* **180**, 61–89 (2010).
19. Havriliak, S. & Negami, S. A complex plane analysis of *α*-dispersions in some polymer systems. *J. Polym. Sci. Part C* **14**, 99–117 (1966).
20. Cole, K. S. & Cole, R. H. Dispersion and absorption in dielectrics I. Alternating current characteristics. *J. Chem. Phys.* **9**, 341–351 (1941).
21. Davidson, D. W. & Cole, R. H. Dielectric relaxation in glycerol, propylene glycol, and n-propanol. *J. Chem. Phys.* **19**, 1484–1490 (1951).
22. Schönhals, A. & Kremer, F. Theory of Dielectric Relaxation, in *Broadband Dielectric Spectroscopy* (ed. Kremer, F. & Schönhals A.) 1–33 (Springer, Berlin, 2002).
23. Jonscher, A. K. *Dielectric Relaxations in Solids* (Chelsea Dielectrics Press, London, 1983).
24. Long, R. Electronic transport in amorphous semiconductors. *Adv. Phys.* **31**, 553–637 (1982).
25. Elliott, S. R. Ac conduction in amorphous chalcogenide and pnictide semiconductors. *Adv. Phys.* **36**, 135–217 (1987).
26. Funke, K. Ion dynamics and correlations. *Philos. Mag. A* **68**, 711–724 (1993).

27. Emmert, S. *et al*. Electrode polarization effects in broadband dielectric spectroscopy. *Eur. Phys. J. B* **83**, 157–165 (2011).
28. Sillescu, H. Heterogeneity at the glass transition: a review. *J. Non-Cryst. Solids* **243**, 81–108 (1999).
29. Ediger, M. D. Spatially heterogeneous dynamics in supercooled liquids. *Annu. Rev. Phys. Chem.* **51**, 99–128 (2000).
30. Richert, R. Heterogeneous dynamics in liquids: fluctuations in space and time. *J. Phys.: Condens. Matter* **14**, R703–R738 (2002).
31. Szabat, B., Weron, K., Hetman, P. Heavy-tail properties of relaxation time distributions underlying the Havriliak–Negami and the Kohlrausch–Williams-Watts relaxation patterns. *J. Non-Cryst. Solids* **353**, 4601–4607 (2007).
32. Boersma, A., van Turnhout, J., Wübbenhorst, M. Dielectric characterization of a thermotropic liquid crystalline copolyesteramide: 1. Relaxation peak assignment. *Macromolecules* **31**, 7453–7460 (1998).
33. Vogel, H. Das Temperaturabhängigkeitsgesetz der Viskositäten von Flüssigkeiten. *Z. Phys.* **22**, 645–646 (1921).
34. Fulcher, G. S. Analysis of recent measurements of viscosity of glasses, *J. Am. Ceram. Soc.* **8**, 339–355 (1925).
35. Tammann, G. & Hesse, W. Die Abhängigkeit der Viskosität von der Temperatur bei unterkühlten Flüssigkeiten. *Z. Anorg. Allg. Chem.* **156**, 245–257 (1926).
36. Angell, C. A. Strong and fragile liquids, in *Relaxations in Complex Systems* (ed. Ngai, K. L. & Wright, G. B.) 3–11 (NRL, Washington, 1985).
37. Ediger, M. D., Angell, C. A. & Nagel, S. R. Supercooled liquids and glasses. *J. Phys. Chem.* **100**, 13200–13212 (1996).
38. Schönhals, A. Molecular Dynamics in Polymer Model Systems, in *Broadband Dielectric Spectroscopy* (ed. Kremer, F. & Schönhals A.) 225–293 (Springer, Berlin, 2002).
39. Schmidt, M. & Maurer F. H. J. Pressure–Volume–Temperature Properties and Free Volume Parameters of PEO/PMMA Blends. *J. Polym. Sci. Part B Polym. Phys.* **36**, 1061–1080 (1997).
40. Faucher, J. A., Koleske, J. V., Santee, E. R., Stratta, J. J. & Wilson, C. W. Glass Transitions of Ethylene Oxide Polymers. *J. Appl. Phys.* **37**, 3962–3964 (1966).
41. Kalfoglou, N. K. Compatibility of poly(ethylene oxide)–poly(vinyl acetate) blends, *J. Polym. Sci., Part B: Polym. Phys.* **20**, 1259-1267 (1982).
42. Pielichowski, K. & Flejtuch, K. Differential scanning calorimetry studies on poly(ethylene glycol) with different molecular weights for thermal energy storage materials. *Polym. Adv. Technol.* **13**, 690–696 (2002).
43. Smith, T. W. *et al*. Dielectric Spectroscopy of Polystyrene/Poly(ethylene oxide) Composites. *Macromolecules* **29**, 5042–5045 (1996).
44. Genix, A. C. *et al*. Dynamics of poly(ethylene oxide) in a blend with poly(methyl methacrylate): A quasielastic neutron scattering and molecular dynamics simulations study. *Phys. Rev. E* **72**, 031808 (2005).
45. Brodeck, M. *et al*. Study of the dynamics of poly (ethylene oxide) by combining molecular dynamic simulations and neutron scattering experiments. *J. Chem. Phys.* **130**, 094908 (2009).
46. Brodeck, M., Alvarez, F., Moreno, A. J., Colmenero, J. & Richter, D. Chain Motion in Nonentangled Dynamically Asymmetric Polymer Blends: Comparison between Atomistic Simulations of PEO/PMMA and a Generic Bead–Spring Model. *Macromolecules* **43**, 3036–3051 (2010).
47. Colmenero, J. Non-exponential Rouse correlators and generalized magnitudes probing chain dynamics. *J. Non-Cryst. Solids* **407**, 302–308 (2015)
48. Stockmayer, W. H. Dielectric dispersion in solutions of flexible polymers. *Pure Appl. Chem.* **15**, 539–554 (1967).
49. Adachi, K. & Kotaka, T. Dielectric normal-mode relaxation. *Prog. Polym. Sci.* **18**, 585–622 (1993).
50. Johari, G. P. & Goldstein, M. Viscous liquids and the glass transition. II. Secondary relaxations in glasses of rigid molecules. *J. Chem. Phys.* **53**, 2372–2388 (1970).
51. Roland, C. M. Relaxation phenomena in vitrifying polymers and molecular liquids. *Macromolecules* **43**, 7875–7890 (2010).
52. Ngai, K. L. & Paluch, M. Classification of secondary relaxation in glass-formers based on dynamic properties. *J. Chem. Phys.* **120**, 857–873 (2004).
53. Stillinger, F. H. A Topographic View of Supercooled Liquids and Glass Formation. *Science* **267**, 1935–1939 (1995).
54. Harmon, J. S., Demetriou, M. D., Johnson, W. L. & Samwer, K. Anelastic to Plastic Transition in Metallic Glass-Forming Liquids. *Phys. Rev. Lett.* **99**, 135502 (2007).
55. Ngai, K. L. Correlation between the secondary β-relaxation time at $T_g$ with the Kohlrausch exponent of the primary α relaxation or the fragility of glass-forming materials. *Phys. Rev. E* **57**, 7346–7349 (1998).

56. Ngai, K. L. Importance of Experiments That Can Test Theories Critically. *J. Phys. Chem. B* **128**, 10709−10726 (2024).
57. Jin, X., Zhang, S. & Runt, J. Broadband Dielectric Investigation of Amorphous Poly(methyl methacrylate)/Poly(ethylene oxide) Blends. *Macromolecules* **37**, 8110−8115 (2004).
58. Böhmer, R., Maglione, M., Lunkenheimer, P. & Loidl, A. Radio-frequency dielectric measurements at temperatures from 10 to 450 K. *J. Appl. Phys.* **65**, 901–904 (1989).